\documentclass[aps,pra,reprint,preprintnumbers,amsmath,amssymb,superscriptaddress,showpacs,showkeys,longbibliography]{revtex4-1}

\usepackage{graphicx}
\usepackage{dcolumn}
\usepackage{bm}
\usepackage{color}
\usepackage{bbold}
\usepackage{braket}
\usepackage{units}
\usepackage{lipsum}
\usepackage{amsmath}
\usepackage{siunitx}
\usepackage{mathtools} 
\usepackage{extarrows}
\usepackage{natbib}

\usepackage[english]{babel}
\usepackage[
	colorlinks=true,
	frenchlinks=false,
        linkcolor=blue,
	    anchorcolor=blue,
        citecolor=blue,
        filecolor=blue,
        urlcolor=blue,
        bookmarks=true,
        bookmarksopen=true,
	    bookmarksnumbered=true,
        bookmarksopenlevel=1,
        plainpages=false,
        pdfpagelabels=true,
	    draft=false]{hyperref}

\DeclareMathOperator*{\argmin}{arg\,min}

\makeatletter
\def\p@subsection{}
\makeatother

\begin{document}

\title[]{Flow of quantum correlations in noisy two-mode squeezed microwave states}

\date{\today}

\author{M. Renger}
\email[]{michael.renger@wmi.badw.de}
\affiliation{Walther-Mei{\ss}ner-Institut, Bayerische Akademie der Wissenschaften, 85748 Garching, Germany}
\affiliation{Physik-Department, Technische Universit\"{a}t M\"{u}nchen, 85748 Garching, Germany}

\author{S. Pogorzalek}
\affiliation{Walther-Mei{\ss}ner-Institut, Bayerische Akademie der Wissenschaften, 85748 Garching, Germany}
\affiliation{Physik-Department, Technische Universit\"{a}t M\"{u}nchen, 85748 Garching, Germany}

\author{F. Fesquet}
\affiliation{Walther-Mei{\ss}ner-Institut, Bayerische Akademie der Wissenschaften, 85748 Garching, Germany}
\affiliation{Physik-Department, Technische Universit\"{a}t M\"{u}nchen, 85748 Garching, Germany}

\author{K. Honasoge}
\affiliation{Walther-Mei{\ss}ner-Institut, Bayerische Akademie der Wissenschaften, 85748 Garching, Germany}
\affiliation{Physik-Department, Technische Universit\"{a}t M\"{u}nchen, 85748 Garching, Germany}

\author{F. Kronowetter}
\affiliation{Walther-Mei{\ss}ner-Institut, Bayerische Akademie der Wissenschaften, 85748 Garching, Germany}
\affiliation{Physik-Department, Technische Universit\"{a}t M\"{u}nchen, 85748 Garching, Germany}
\affiliation{Rohde \& Schwarz GmbH \& Co. KG, M\"{u}hldorfstra{\ss}e 15, 81671 Munich, Germany}

\author{Q. Chen}
\affiliation{Walther-Mei{\ss}ner-Institut, Bayerische Akademie der Wissenschaften, 85748 Garching, Germany}
\affiliation{Physik-Department, Technische Universit\"{a}t M\"{u}nchen, 85748 Garching, Germany}

\author{Y. Nojiri}
\affiliation{Walther-Mei{\ss}ner-Institut, Bayerische Akademie der Wissenschaften, 85748 Garching, Germany}
\affiliation{Physik-Department, Technische Universit\"{a}t M\"{u}nchen, 85748 Garching, Germany}

\author{K. Inomata}
\affiliation{RIKEN Center for Quantum Computing (RQC), Wako, Saitama 351-0198, Japan}
\affiliation{National Institute of Advanced Industrial Science and Technology, 1-1-1 Umezono, Tsukuba, Ibaraki, 305-8563, Japan}

\author{Y. Nakamura}
\affiliation{RIKEN Center for Quantum Computing (RQC), Wako, Saitama 351-0198, Japan}
\affiliation{Department of Applied Physics, Graduate School of Engineering, The University of Tokyo, Bunkyo-ku, Tokyo 113-8656, Japan}

\author{A. Marx}
\affiliation{Walther-Mei{\ss}ner-Institut, Bayerische Akademie der Wissenschaften, 85748 Garching, Germany}

\author{F. Deppe}
\affiliation{Walther-Mei{\ss}ner-Institut, Bayerische Akademie der Wissenschaften, 85748 Garching, Germany}
\affiliation{Physik-Department, Technische Universit\"{a}t M\"{u}nchen, 85748 Garching, Germany}
\affiliation{Munich Center for Quantum Science and Technology (MCQST), Schellingstr. 4, 80799 Munich, Germany}

\author{R. Gross}
\email[]{rudolf.gross@wmi.badw.de}
\affiliation{Walther-Mei{\ss}ner-Institut, Bayerische Akademie der Wissenschaften, 85748 Garching, Germany}
\affiliation{Physik-Department, Technische Universit\"{a}t M\"{u}nchen, 85748 Garching, Germany}
\affiliation{Munich Center for Quantum Science and Technology (MCQST), Schellingstr. 4, 80799 Munich, Germany}

\author{K. G. Fedorov}
\email[]{kirill.fedorov@wmi.badw.de}
\affiliation{Walther-Mei{\ss}ner-Institut, Bayerische Akademie der Wissenschaften, 85748 Garching, Germany}
\affiliation{Physik-Department, Technische Universit\"{a}t M\"{u}nchen, 85748 Garching, Germany}

\begin{abstract}
We study nonclassical correlations in propagating two-mode squeezed microwave states in the presence of noise. We focus on two different types of correlations, namely, quantum entanglement and quantum discord. Quantum discord has various intriguing fundamental properties which require experimental verification, such as the asymptotic robustness to environmental noise. Here, we experimentally investigate quantum discord in propagating two-mode squeezed microwave states generated via superconducting Josephson parametric amplifiers. By exploiting an asymmetric noise injection into these entangled states, we demonstrate the robustness of quantum discord against thermal noise while verifying the sudden death of entanglement. Furthermore, we investigate the difference between quantum discord and entanglement of formation, which can be directly related to the flow of locally inaccessible information between the environment and the bipartite subsystem. We observe a crossover behavior between quantum discord and entanglement for low noise photon numbers, which is a result of the tripartite nature of noise injection. We demonstrate that the difference between entanglement and quantum discord can be related to the security of certain quantum key distribution protocols.
\end{abstract}

\maketitle
\section{Introduction}
Quantum communication and quantum computing protocols often employ entanglement as a nonclassical resource to provide improvement of information transfer and achieve a quantum speed-up in processing \cite{Braunstein2005}. In this regard, entanglement plays a key role in realizing quantum error correction \cite{Chen_2021}, efficient quantum simulation \cite{Eddins_2022}, as well as in achieving quantum supremacy \cite{Arute_2019}. Prominent examples of quantum communication protocols are quantum key distribution \cite{Wang2007}, dense coding \cite{Braunstein2000}, or quantum teleportation \cite{Fedorov2021, Furusawa1998}, where entanglement is exploited for an efficient and unconditionally secure state transfer. However, entanglement represents only one particular quantum resource and does not capture all nonclassical correlations. In particular, quantum discord (QD) provides a more general measure of nonclassical correlations including entanglement \cite{Ollivier2001}. Quantum discord may serve as a resource in multiple quantum information processing protocols such as deterministic quantum computing with one qubit (DQC1) \cite{Caves2008,Lanyon2008}, quantum sensing \cite{Adesso2014}, and quantum illumination \cite{Weedbrook2016}. Furthermore, theoretical investigations imply that QD has multiple intriguing physical properties which still lack experimental verification, such as the asymptotic robustness against environmental noise and its relation to entanglement in mixed tripartite systems \cite{Weedbrook2012}. These two properties are the key focus of our experimental study. In particular, our experiments reveal that the process of noise injection into a bipartite quantum system necessarily creates multipartite quantum correlations with the environment. As a result, the noise suppresses quantum correlations in the bipartite system and simultaneously increases the correlations between one of the subsystems and the environment. We measure this effect by extracting a flow of locally inaccessible information (LII) \cite{Fanchini2012}. This LII flow behaves fundamentally different as a function of noise for different subsystems. The more detailed investigation of the LII flow may be relevant for testing ideas related to quantum Darwinism \cite{Girolami_2022, Zurek_2009}, where correlations with different fragments of the environment eventually lead to an objective reality \cite{Touil_2022}. In addition, we demonstrate that LII is related to the unconditional security of a certain class of quantum key distribution (QKD) protocols.

The potential robustness of QD versus noise is extremely useful for various quantum communication protocols, where unavoidable external fluctuations cause loss of quantum correlations, thereby lowering the efficiency of these protocols. Protocols based on quantum entanglement are particularly vulnerable to noise, since entanglement cannot survive significant noise levels, leading to the so-called sudden death of entanglement \cite{Yu2009}. This is especially important for quantum communication protocols in the microwave regime, as room temperature is associated with thousands of thermal noise photons with characteristic frequencies of several GHz \cite{Sanz2018}. Here, the expected asymptotic robustness of QD \cite{Werlang2009} against noise offers a natural path for quantum communication and sensing. An actual challenge is to find protocols capable of exploiting QD as a quantum resource. As such, the remote state preparation (RSP) protocol stands out as one of the prominent known examples. RSP aims at the generation of a desired and known quantum state at a remote location with the assistance of classical communication and complementary nonclassical correlations \cite{Bennet2001}. A finite quantum advantage provided by RSP is associated with the smaller amount of classical information required to prepare a quantum state, as compared to a fully classical communication protocol, and the unconditional security of the feedforward signal \cite{Pogorzalek2019}. In some scenarios, the RSP protocol appears to exploit QD as its nonclassical resource \cite{Dakic2012, Ollivier2001}. Finally, there is a class of quantum sensing protocols known as quantum illumination, where entangled light is used to detect the presence of a low-reflectivity object in a bright noisy background. Naturally, entanglement vanishes in this scenario, yet residual non-classical correlations, captured by QD, persist. The latter seems to be connected to the quantum advantage of these quantum sensing schemes \cite{Lloyd2008,Weedbrook2016}.

\begin{figure}
	\begin{center}
		\includegraphics[width=\linewidth,angle=0,clip]{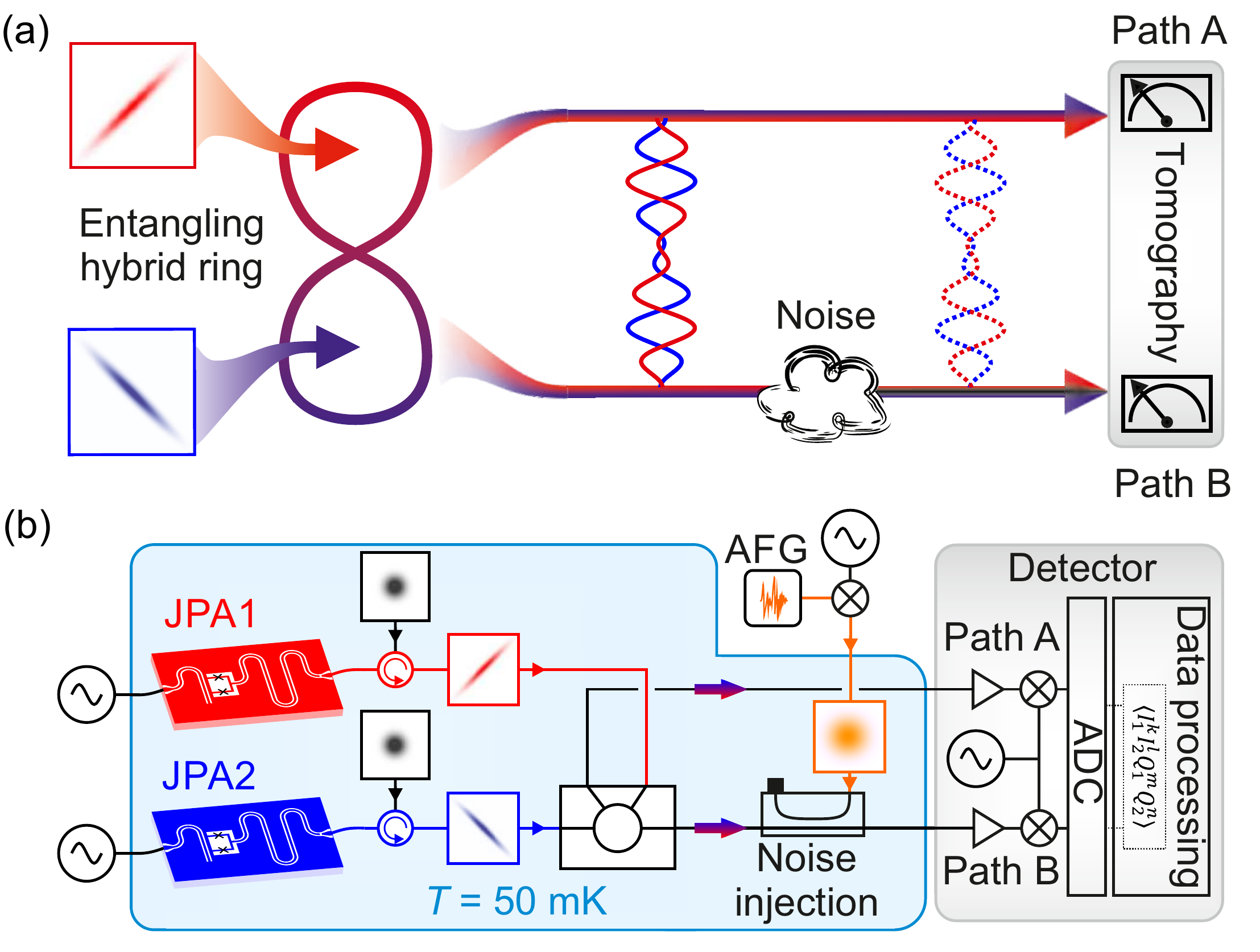}
	\end{center}
	\caption{(a) Scheme of two-mode squeezing and noise injection. Nonlocal correlations are generated by superposing two orthogonally squeezed states on a symmetric beam splitter to generate a path-entangled frequency-degenerate TMS state. The asymmetric noise injection couples the environment to one of the TMS subsystems. (b) Equivalent experimental setup consisting of two JPAs for squeezed state generation, a microwave hybrid ring, and a directional coupler for injection of white Gaussian noise generated at room temperature by an arbitrary function generator (AFG). The two-mode signal is detected with a heterodyne receiver setup and digitally processed to extract the statistical signal moments and reconstruct the corresponding covariance matrix.}
	\label{Fig:Fig1}
\end{figure}

\begin{figure*}
	\begin{center}
		\includegraphics[width=\linewidth,angle=0,clip]{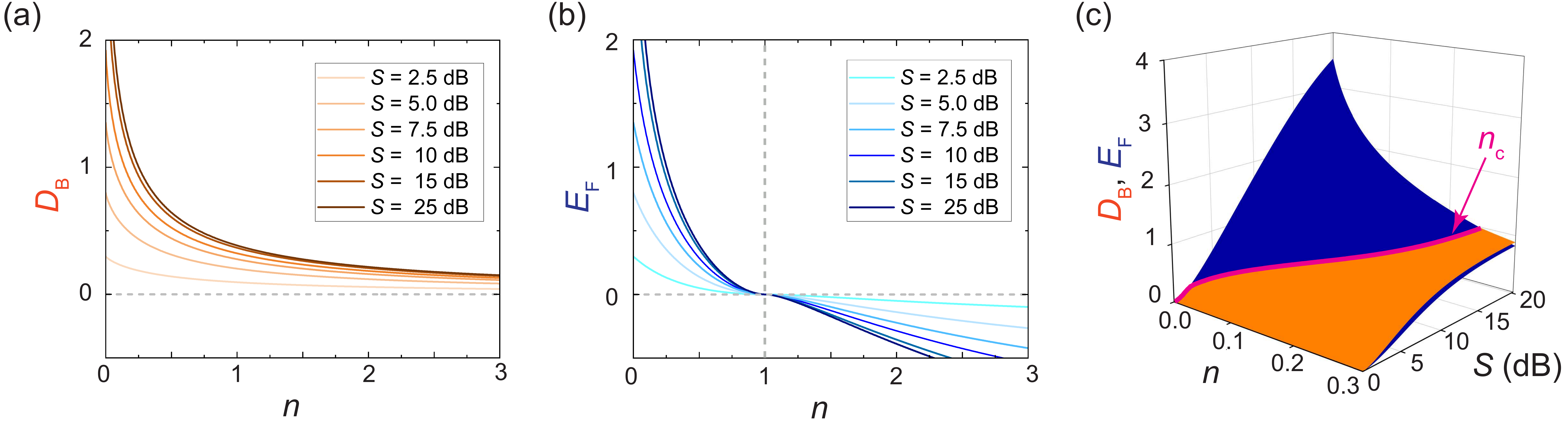}
	\end{center}
	\caption{Theoretical prediction of quantum correlations as a function of squeezing level $S$ and noise photon number $n$. Panel~(a) shows the quantum discord $D_\mathrm{B}$, thereby demonstrating the asymptotic robustness, $D_\mathrm{B} > 0$, for any finite level of noise and squeezing. Panel (b) shows the analytical lower bound $E_\mathrm{F}$ for EoF. At $n = 1$, as indicated by the grey dashed line, we observe $E_\mathrm{F} = 0$, independent of $S$, which demonstrates the sudden death of entanglement. In panel (c), $E_\mathrm{F}$ and $D_\mathrm{B}$ are plotted in the regime of $n \ll 1$, revealing a crossover region. Here, for low $n$, EoF is larger than QD. For increasing $n$,  QD becomes larger than EoF. The crossover noise photon number $n_\mathrm{c}$ is depicted in magenta.}
	\label{Fig:Fig2}
\end{figure*}

\section{Experiment}
In this paper, we experimentally study the effect of noise injection into one mode of a propagating two-mode squeezed (TMS) state which is distributed along two paths, A and B. A schematic illustration of this scenario is shown in Fig.\,\ref{Fig:Fig1}(a). The TMS state is generated by superimposing two orthogonally squeezed microwave states at a symmetric microwave beam splitter. Then, we inject an uncorrelated broadband noise into path B. Finally, we perform a joint quantum state tomography, which allows us to extract full information about the two-mode quantum state \cite{Ralph2017,Paris2010}. We use the entanglement of formation (EoF), $\mathcal{E}_\mathrm{F}$, as a measure for bipartite entanglement between parties A and B. We choose this specific entanglement measure since it exactly coincides with QD for pure states \cite{Adesso2010}. Simultaneously, EoF is directly related to QD via various monogamy relations, which eventually enables the connection of both quantities to LII \cite{Fanchini2011, Fanchini2012}. For continuous-variable quantum systems, EoF quantifies the minimal amount of two-mode squeezing needed to prepare an entangled state, starting from a classical one by using local operations and classical communication \cite{Tserkis2019}. In addition, we extract an asymmetric bipartite QD, $D_\mathrm{A}$ ($D_\mathrm{B}$), between the two subsystems A and B. It is defined as the difference between their quantum mutual information $I_{\mathrm{AB}}$ and a one-way classical correlation $J_{\mathrm{A}|\mathrm{B}}$ ($J_{\mathrm{B}|\mathrm{A}}$),
\begin{equation}
    D_\mathrm{A} = I_{\mathrm{AB}} - J_{\mathrm{A}|\mathrm{B}}, \qquad D_\mathrm{B} = I_{\mathrm{AB}} - J_{\mathrm{B}|\mathrm{A}},
\end{equation}
and quantifies the non-local fraction of $I_{\mathrm{AB}}$  \cite{Henderson2001}.

Figure\,\ref{Fig:Fig1}(b) illustrates our experimental setup. We use two superconducting flux-driven Josephson parametric amplifiers (JPAs) operated at the frequency $\omega_0/2\pi = \SI{5.323}{\giga \hertz}$ for squeezed state generation, where each JPA performs a squeezing operation on the incident weak thermal state \cite{Yamamoto2008}. By sending the respective squeezed states to a hybrid ring (symmetric microwave beam splitter), we \mbox{generate} a TMS microwave state. Here, entanglement is expressed in strong correlations between two nonlocal field quadratures \cite{Menzel2012,Fedorov2016,Fedorov2018,Flurin2012}. Our final state tomography is based on heterodyne measurements of paths A and B. After digital down-conversion and filtering, we extract the statistical field quadrature moments using a reference state reconstruction method \cite{Menzel2012, Eichler2011}. Under the assumption that the reconstructed states are Gaussian, a local phase space distribution is described by the resulting two-mode covariance matrix \cite{Menzel2012,Fedorov2016,Fedorov2018}. The photon number calibration of the experimental setup is obtained by using Planck spectroscopy \cite{Mariantoni2010}. In order to test the robustness of the nonclassical correlations against noise, we perform a controlled noise injection into one of the entangled paths \cite{Menzel2012,Fedorov2016,Fedorov2018}. The noise signal is generated using an arbitrary function generator (AFG) which produces a low-frequency white Gaussian noise with a specified bandwidth of $\SI{160}{\mega \hertz}$. This noise signal is upconverted to the carrier frequency $\omega_0/2\pi$ and guided into the cryogenic setup. We implement the actual noise injection in one of the entangled paths with a directional coupler with coupling $\beta = \SI{-20}{\decibel}$. By varying the noise power emitted from the AFG, we probe both EoF and QD as a function of the injected noise photon number $n$ for different JPA squeezing levels. The latter is defined as $S = -10\log_{10} \! \left(v_\mathrm{s}/0.25 \right)$, where $v_\mathrm{s}$ is the variance of the squeezed quadrature and the chosen vacuum reference is $v_\mathrm{vac} = 0.25$. More details about the experimental setup are provided in Appendix\,\ref{sec:setup}.

\section{Entanglement of formation and quantum discord}
Under the assumption that the state is Gaussian, EoF and QD can be extracted from the reconstructed two-mode covariance matrix $V_{\mathrm{AB}}$ \cite{Paris2010, Ralph2020}. In Ref.\,\citenum{Ralph2017}, it has been shown that a lower bound, $E_\mathrm{F}$, for Gaussian EoF can be expressed as
\begin{equation}\label{Eq:lower bound EoF}
        E_\mathrm{F} = s_\gamma \left[\cosh^2 \! \gamma \ln  (\cosh^2 \! \gamma) - \sinh^2 \! \gamma \ln  (\sinh^2 \! \gamma) \right] \leq \mathcal{E}_\mathrm{F},
\end{equation}
where $\gamma$ represents the minimally required amount of two-mode squeezing to disentangle the respective bipartite quantum state and $s_\gamma = \mathrm{sign} (\gamma)$. For our analysis, we use an approximation $E_\mathrm{F} \simeq \mathcal{E_\mathrm{F}}$, which becomes exact in the case of symmetric local states A and B. The asymmetric Gaussian QD, $D_\mathrm{A}$, corresponds to the correlation left after we perform a local measurement on subsystem~B. It can be calculated as
\begin{equation}\label{Eq.:DB}
    D_\mathrm{A} = f\left( \sqrt{I_2} \right) - f(\nu_+) - f(\nu_-) + f\left( \sqrt{E_{\mathrm{A}|\mathrm{B}}^{\min}}\,\right),
\end{equation}
where $I_2$ denotes the second symplectic invariant of $V_{\mathrm{AB}}$ and $\nu_{\pm}$ are the corresponsing symplectic eigenvalues \cite{Adesso2010}. The quantity $\sqrt{E_{\mathrm{A}|\mathrm{B}}^{\min}}$ describes the minimized conditional entropy and $f$ is defined as $f(x) = \left(2x + \frac{1}{2} \right)\ln \! \left(2x + \frac{1}{2} \right) - \left(2x - \frac{1}{2} \right) \ln \! \left(2x - \frac{1}{2} \right)$. A similar expression can be written for the quantum discord $D_\mathrm{B}$, where the measurement is performed on system A. It can be shown that for pure quantum systems, i.e., in the limit of $n \to 0$, EoF and QD coincide \cite{Adesso2010}. Nevertheless, for mixed states, these quantities behave fundamentally different, as theory predicts asymptotic stability of QD, in contrast to EoF.

Figure \ref{Fig:Fig2} shows theoretically expected results for an idealized experiment with zero losses and noiseless JPAs. In Fig.\,\ref{Fig:Fig2}(a) [Fig.\,\ref{Fig:Fig2}(b)], we plot $D_\mathrm{B}$ ($E_\mathrm{F}$) as a function of the average noise photon number $n$, injected to the TMS state, and the squeezing level $S$. We observe the expected asymptotic stability of QD, $D_\mathrm{B} > 0$, and the sudden death of entanglement, $E_\mathrm{F} < 0$, for $n > 1$. The latter effect can be understood by expressing $\gamma$ in Eq.\,\ref{Eq:lower bound EoF} analytically as
\begin{equation}\label{Eq:gamma}
     \gamma(r, n) = \frac{1}{2} \ln \left[\frac{e^{2r} + n}{1 + e^{2r}n}\right],
\end{equation}
where $r$ is the squeezing factor, which can be calculated by $r = S/(20\log_{10} e)$ for noiseless amplification by both JPAs. We observe that $\gamma (r, 1) = 0$, independent of $r$, indicating the sudden death of entanglement. In Fig.\,\ref{Fig:Fig2}(c), we plot the theory values of $E_\mathrm{F}$ and $D_\mathrm{B}$ for low noise photon numbers and observe a crossover between EoF and QD. We denote the crossover point in terms of a corresponding noise photon number, $n_\mathrm{c}$. This crossover point, $n_\mathrm{c} > 0$, exists for any positive squeezing level $S$. This effect has been predicted in Refs.\,\citenum{Luo2008} and  \citenum{Marian2015}, and is a direct result of a tripartite nature of the noise injection. The latter implies that a correct quantum mechanical description of noise necessarily requires to take into account the environment E as a third interacting quantum system. For the tripartite system, it can be shown that EoF and QD are monogamous, i.e., that bipartite QD can only be increased by the simultaneous consumption of bipartite EoF and vice versa \cite{Fanchini2011}. From this monogamic conservation relation, it has been shown that the difference between EoF and QD can be expressed as \cite{Fanchini2012}
\begin{align}\label{Eq: LLI flow A}
    & \Delta_\mathrm{A} \equiv D_\mathrm{A} - \mathcal{E}_\mathrm{F} = \frac{1}{2}\left(\mathcal{L}_{\mathrm{B} \to \mathrm{A} \to \mathrm{E}} - \mathcal{L}_{\mathrm{E} \to \mathrm{A} \to \mathrm{B}} \right), \\ \label{Eq: LLI flow B}
    & \Delta_\mathrm{B} \equiv D_\mathrm{B} - \mathcal{E}_\mathrm{F} = \frac{1}{2}\left(\mathcal{L}_{\mathrm{A} \to \mathrm{B} \to \mathrm{E}} - \mathcal{L}_{\mathrm{E} \to \mathrm{B} \to \mathrm{A}} \right), \\ \label{Eq: LLI flow AB}
    & \Delta_\mathrm{AB} \equiv \frac{D_\mathrm{A} + D_\mathrm{B}}{2} - \mathcal{E}_\mathrm{F} = \frac{1}{2}\left(\mathcal{L}_{\binom{\mathrm{A}}{\mathrm{B}} \to \mathrm{E}} - \mathcal{L}_{\mathrm{E} \to \binom{\mathrm{A}}{\mathrm{B}}} \right),
\end{align}
where $\mathcal{L}_{X \to Y \to Z}$ denotes the flow of LII from the system $X$ over $Y$ to $Z$ and $\mathcal{L}_{\binom{X}{Y} \to Z}$ ($\mathcal{L}_{Z \to \binom{X}{Y} }$) is the LII flow from (to) the bipartite system $XY$ to (from) $Z$. As a result, if $\Delta_\mathrm{AB} > 0$, more LII flows from the bipartite system $\mathrm{AB}$ to the environment E than vice versa.

\begin{figure*}
	\begin{center}
		\includegraphics[width=0.9\linewidth,angle=0,clip]{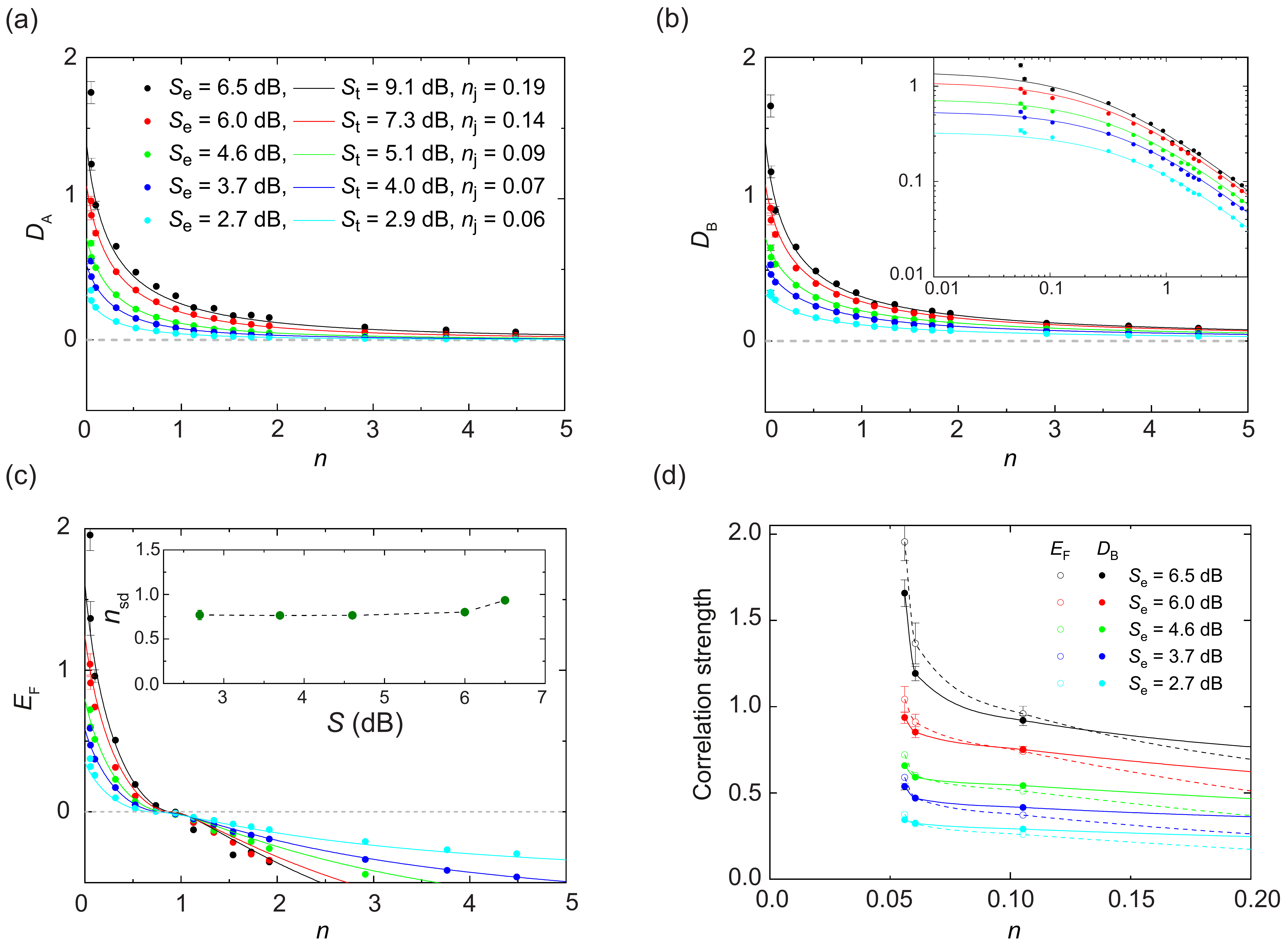}
	\end{center}
	\caption{(a) Experimentally obtained values of quantum discord $D_\mathrm{A}$ as a function of the injected noise photon number $n$ for various squeezing levels $S$. Dots indicate the measured data and lines are fits according to a realistic model, described in Appendix\,\ref{Sec:fit}, which takes a finite JPA noise into account. The quantity $S_\mathrm{e}$ denotes the experimentally determined squeezing level and $S_\mathrm{t}$ is the corresponding squeezing level, obtained by fitting the data by the theory prediction. The JPA noise $n_\mathrm{j}$ is extracted from the fit and is a function of gain. Although only shown for $D_\mathrm{A}$, the fitted values for $S_\mathrm{e}$ and $n_\mathrm{j}$ are the same for $D_\mathrm{B}$ and $E_\mathrm{F}$. (b) Experimentally obtained values of quantum discord $D_\mathrm{B}$ as a function of the injected noise photon number, $n$, for various squeezing levels. The inset shows the same data in a log-log plot. (c) Experimental EoF (full circles) and corresponding fits (lines) for various squeezing levels. We observe the sudden death of entanglement at $n_\mathrm{sd} \simeq 1$. The inset shows that $n_\mathrm{sd}$ is independent of $S$, where $n_\mathrm{sd}$ is obtained from the experimental data using cubic Hermite spline interpolation. Error bars are obtained from the statistical measurement error and are only plotted if the error exceeds the symbol size. (d) Zoom-in of experimental results for $D_\mathrm{B}$ and $E_\mathrm{F}$ for low noise photon numbers $n$ and various squeezing levels $S_\mathrm{e}$. Solid (dashed) lines are the result from a cubic Hermite spline interpolation between the measured values for $D_\mathrm{B}$ (EoF). Here, we observe the crossover behavior of QD and EoF, as predicted by the theory.}
	\label{Fig:Fig3}
\end{figure*}

\begin{figure*}
	\begin{center}
		\includegraphics[width=0.9\linewidth,angle=0,clip]{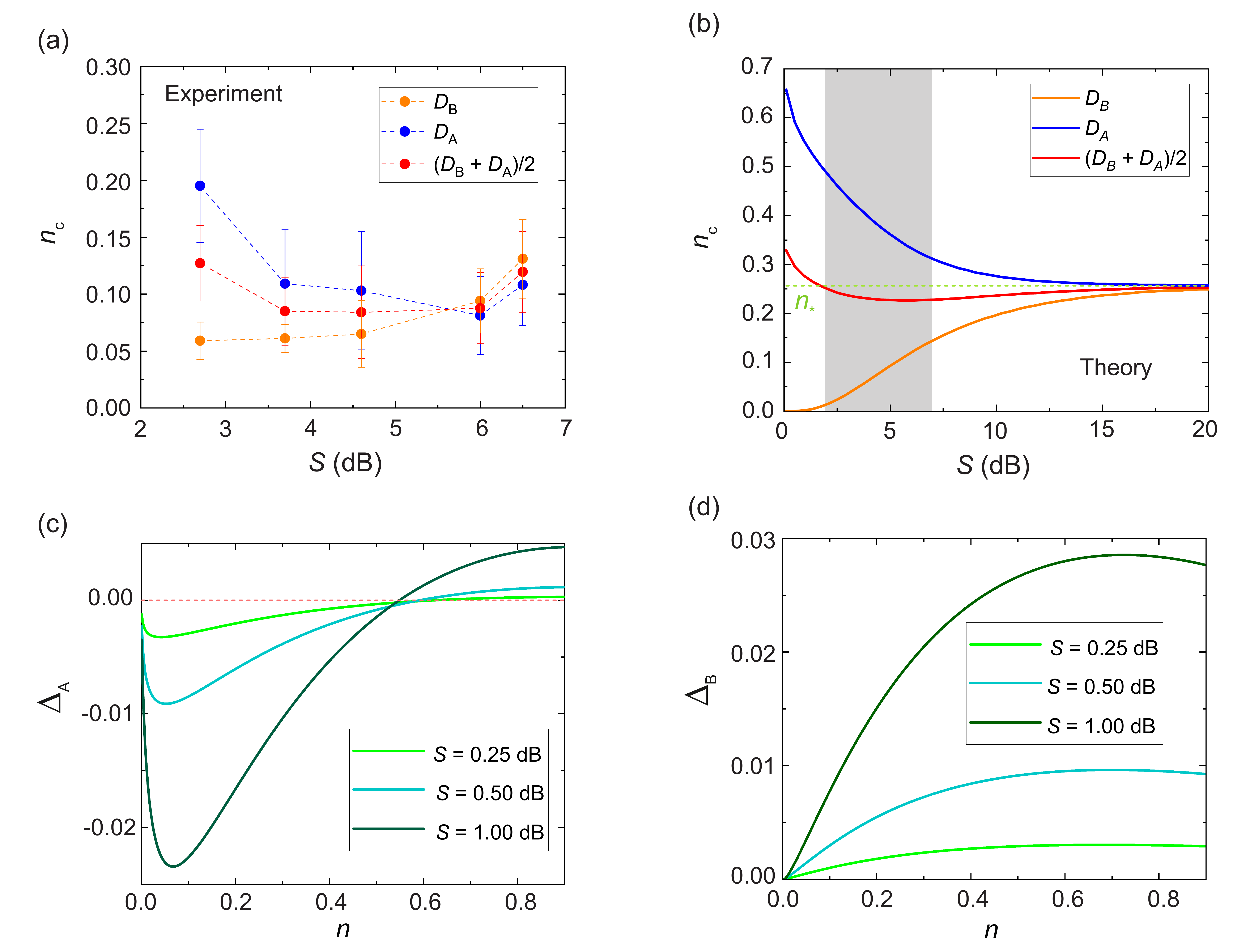}
	\end{center}
	\caption{(a) Experimentally determined crossover noise photon number, $n_\mathrm{c}$, as a function of the squeezing level for $D_\mathrm{A}$ (blue) and $D_\mathrm{B}$ (orange). Red dots represent the arithmetic mean of $n_\mathrm{c}$ for $D_\mathrm{A}$ and $D_\mathrm{B}$, where we observe a minimum in the region of $\SI{5}{\decibel}$. The error bars are determined from the experimental uncertainties of QD and EoF by randomized error sampling. (b) Theoretically predicted $n_\mathrm{c}$ for $D_\mathrm{A}$ (blue), $D_\mathrm{B}$ (orange) for the case of an idealized (lossless and noiseless, $\beta \to 0$) experiment. In the case of $D_\mathrm{A}$ ($D_\mathrm{B}$),  $n_\mathrm{c}$ decreases (increases) monotonically with $S$. We observe a minimum for $S_\mathrm{min} \simeq \SI{5.73}{\decibel}$ for the red curve which corresponds to the arithmetic mean of both discords, $D_\mathrm{A}$ and $D_\mathrm{B}$. In the limit $S \to \infty$, $n_\mathrm{c}$ converges to the same constant $n_* \simeq 0.26$ for $D_\mathrm{A}$ and $D_\mathrm{B}$. The region of experimentally obtained squeezing levels, corresponding to panel (a), is indicated in gray. (c) Theoretical difference $\Delta_\mathrm{A}$ between $D_\mathrm{A}$ and $E_\mathrm{F}$ as a function of the noise photon number $n$ for various squeezing levels $S$. (d) Theoretical difference $\Delta_\mathrm{B}$ between $D_\mathrm{B}$ and $E_\mathrm{F}$ as a function of the noise $n$ for various squeezing levels $S$.}
	\label{Fig:Fig4}
\end{figure*}
\begin{figure}[!h]
	\begin{center}
		\includegraphics[width=0.75\linewidth,angle=0,clip]{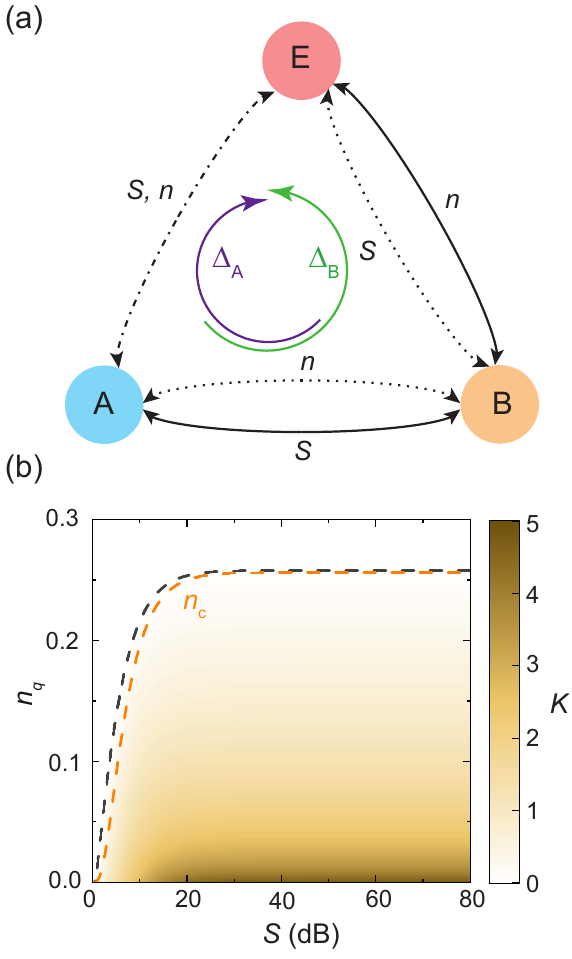}
	\end{center}
	\caption{(a) Schematic dependence of bipartite correlations between subsystems A, B, and E. Solid (dotted) black arrows indicate a monotonic increase (decrease) of bipartite correlations with a respective quantity. The dotted-dashed arrow, connecting A and E, indicates that correlations \mbox{between} these subsystems are not necessarily monotonic in $S$ or $n$. The curved purple (green) arrow indicates the LII flow $\mathrm{B}{\to}\mathrm{A}{\to}\mathrm{E}$ ($\mathrm{A}{\to}\mathrm{B}{\to}\mathrm{E}$), described by $\Delta_\mathrm{A}$ ($\Delta_\mathrm{B}$). (b) Theoretical secret key $K$ as a function of the resource state squeezing level $S$ and injected noise photon number $n_q$ in the detected quadrature. Here, we assume a continuous-variable QKD protocol between A and B \cite{Cerf2001}, where the environment acts as an eavesdropper. The black dashed line indicates the threshold separating the areas of positive secret keys (secure), $K > 0$, and negative keys (insecure), $K <  0$. The orange dashed line shows the corresponding $n_\mathrm{c}$ for $D_\mathrm{B}$ as a function of $S$, which offers an intuitive explanation for the security of the QKD protocol on the language of the LII flow between the subsystems.}
	\label{Fig:Fig5}
\end{figure}

\section{Results and discussion}
The experimentally determined QD values, $D_\mathrm{A}$ and $D_\mathrm{B}$, are provided in Fig.\,\ref{Fig:Fig3}(a) and Fig.\,\ref{Fig:Fig3}(b) and the quantum entanglement measure $E_\mathrm{F}$ is shown in Fig.\,\ref{Fig:Fig3}(c). The line plots correspond to a fit according to Eq.\,\ref{Eq.:DB} for the QD and Eq.\,\ref{Eq:lower bound EoF} for the EoF, where we take the finite JPA noise into account. The gain-dependent noise, added by the JPAs, is modelled by a power law dependence, $n_\mathrm{j} = \chi_1 (G - 1)^{\chi_2}$, where $G$ represents the degenerate gain~\cite{Renger2021}. The coefficients $\chi_1$ and $\chi_2$ are treated as fit parameters. We find $\chi_1 = 0.05$ and $\chi_2 = 0.56$. In Fig.\,\ref{Fig:Fig3}(a), we additionally show the experimentally determined squeezing level $S_\mathrm{e}$, as well as the respective theoretical squeezing level $S_\mathrm{t}$ and fitted JPA noise $n_\mathrm{j}$. We observe that the fits reliably reproduce the experimental data. More information about the fitting routine is given in Appendix\,\ref{Sec:fit}.

Furthermore, we find that the experimentally determined QD is always positive and converges towards zero for $n \to \infty$, thereby, proving the asymptotic robustness against noise. In contrast to that, we find that $E_\mathrm{F}$ becomes zero already at a finite noise photon number $n_\mathrm{sd} \simeq 1$, experimentally verifying the sudden death of entanglement. This value is an important fundamental noise threshold for two-mode squeezed light. The respective experimental values for $n_\mathrm{sd}$ as a function of squeezing are shown in the inset of Fig.\,\ref{Fig:Fig3}(c). They have been extracted from the experimental data using cubic Hermite spline interpolation. We find that $n_\mathrm{sd}$ is independent of the squeezing level, as expected from theory. Note that the experimentally determined noise level for the sudden death of entanglement is lower than the theoretically predicted noise photon number of unity, which is a result from the finite noise added by the JPAs themselves. In addition to that, further deviations from ideal theory are caused by path losses and a pump crosstalk between the JPAs. However, these imperfections are not taken into account in the current fit model. 

For most of the observed states, EoF appears to be smaller than QD. This can be understood by the fact that, by definition, QD describes more general nonlocal correlations than EoF. However, this simple relation is only true in the bipartite limit. When one considers the environment as a third party, the relation between bipartite QD and EoF may change. In order to experimentally investigate the latter we investigate the regime of $n \ll 1$ in more detail. Here, theory predicts a crossover between EoF and QD, according to Fig.\,\ref{Fig:Fig2}(c). To experimentally study this crossover region, we replot the measured $D_\mathrm{B}$ and $E_\mathrm{F}$ for noise photon numbers $n \leq 0.2$ in Fig.\,\ref{Fig:Fig3}(d), revealing the intersection between EoF and QD, especially well observable for $S_\mathrm{e} = \SI{6.5}{\decibel}$. The solid (dashed) lines correspond to a cubic Hermite spline interpolation for $D_\mathrm{B}$ ($E_\mathrm{F}$). From this interpolation, we determine the crossover noise photon number $n_\mathrm{c}$ as a function of squeezing. The same procedure is repeated for $D_\mathrm{A}$ and $D_\mathrm{AB} \equiv (D_\mathrm{A} + D_\mathrm{B})/2$. The corresponding results are plotted in Fig.\,\ref{Fig:Fig4}(a), and the predictions based on an ideal (lossless and noiseless) model are depicted in Fig.\,\ref{Fig:Fig4}(b). In the limit $S \to \infty$, we observe that $n_\mathrm{c} \to n_* \simeq 0.26$ for $D_\mathrm{A}$ as well as for $D_\mathrm{B}$. Furthermore, we find a qualitative agreement between experiment and theory for the dependence of $n_\mathrm{c}$ on the squeezing level $S$. Nevertheless, we observe that for $D_\mathrm{A}$ the experimentally determined values are lower than those predicted by theory. This deviation can be explained by the finite noise, losses, and crosstalk between the JPAs, since these effects are not taken into account in the ideal theory model. Furthermore, we note that we cannot experimentally investigate the whole range shown in Fig.\,\ref{Fig:Fig4}(b), as larger squeezing levels are not experimentally achievable due to the gain-dependent noise added by the JPAs. As shown by the red line in Fig.\,\ref{Fig:Fig4}(b), the corresponding $n_\mathrm{c}$ for $(D_\mathrm{A} + D_\mathrm{B})/2$ has a minimum at $n_\mathrm{min} \simeq 0.23$, corresponding to $S_\mathrm{min} = \SI{5.7}{\decibel}$. Thus, when one attempts to maximize $n_\mathrm{c}$ in the bipartite system AB, it is not always beneficial to increase the squeezing level. Experimental data in Fig.\,\ref{Fig:Fig4}(a) qualitatively reproduces this result.

Next, we investigate the asymmetric differences between EoF and QD, $\Delta_\mathrm{A}$, $\Delta_\mathrm{B}$, and $\Delta_\mathrm{AB}$, as a function of the noise photon number $n$. Figure\,\ref{Fig:Fig4}(c) [(d)] shows the theoretically expected noise dependence of $\Delta_\mathrm{A}$ ($\Delta_\mathrm{B}$) for various squeezing levels $S$. We observe that the quantities $\Delta_\mathrm{A}$ and $\Delta_\mathrm{B}$ behave fundamentally different in the limit of low noise. In particular, the crossover noise photon number $n_\mathrm{c}$ decreases monotonically with increasing $S$ in the case of $\Delta_\mathrm{A}$, as can also be seen in Fig.\,\ref{Fig:Fig4}(b). On the contrary, $n_\mathrm{c}$ shows a monotonic increase as a function of $S$ for increasing $\Delta_\mathrm{B}$. This fundamental deviation can be understood by the fact that noise injection in B is a local process and directly leads to bipartite correlations between party B and environment E and only indirectly correlates A and E. Thus, the bipartite correlations between B and E increase monotonically with $n$. Furthermore, the correlation between A and B monotonically decreases as a function of $n$. In contrast to that, correlations between A and E can only result from an interplay between squeezing $S$ and noise $n$, and are not necessarily required to be monotonic in these quantities. The scenario is schematically depicted in Fig.\,\ref{Fig:Fig5}(a), where solid black arrows, connecting to systems $X$ and $Y$ with $X, Y \in \{\mathrm{A}, \mathrm{B}, \mathrm{E}\}$, indicate a monotonic increase of correlations. The direction of the LII flow, described by $\Delta_\mathrm{A}$ ($\Delta_\mathrm{B}$) according to Eq.\,\ref{Eq: LLI flow A} (Eq.\,\ref{Eq: LLI flow B}), is shown by the purple (green) curved arrow. Note that $\Delta_\mathrm{A}$ describes the net LII flow $\mathrm{B} \to \mathrm{A} \to \mathrm{E}$ and $\Delta_\mathrm{B}$ describes the respective net flow $\mathrm{A} \to \mathrm{B} \to \mathrm{E}$. Therefore, the fundamentally different behavior of these quantities as a function of $n$, as shown in Figs.~\ref{Fig:Fig4}(c) and (d), can be explained by the fact that in the case of $\Delta_\mathrm{B}$ no direct bipartite correlations between A and E are required to establish an LII flow, in contrast to the case described by $\Delta_\mathrm{A}$. 

Furthermore, the quantities $\Delta_\mathrm{A}$, $\Delta_\mathrm{B}$, and $\Delta_\mathrm{AB}$ can become of practical interest for entanglement-based quantum key distribution (QKD) protocols \cite{Usenko2014}, where an eavesdropper attempts to extract LII from a bipartite quantum system. In such a scenario, the subsystems A and B exploit quantum correlations to securely share a common secret key and the subsystem E can be related to an eavesdropper controlling the environment \cite{Ahonen2008}. Then, the noise injection can be interpreted as the result of an entangling cloner attack performed by the eavesdropper~\cite{Grosshans2003}. It directly follows from Eq.\,\ref{Eq: LLI flow AB} that the eavesdropper needs to add at least $n_\mathrm{c}$ noise photons to the system $\mathrm{AB}$ to get a positive net flow of LII. Numerically, we find that in the limit $S \to \infty$, we have $n_\mathrm{c} \to n_* \simeq 0.26$ for $\Delta_\mathrm{A}$, $\Delta_\mathrm{B}$, and $\Delta_\mathrm{AB}$. In order to investigate this interrelation, we consider a Gaussian CV-QKD scheme described in Refs.\,\citenum{Cerf2001} and \citenum{Fesquet2022} under the assumption of reverse reconciliation \cite{Laudenbach2018}. We consider a scenario, where A and B share a TMS resource state, and assume that the eavesdropper performs an entangling cloner attack via the directional coupler. The resulting secret key which quantifies the amount of exchanged secure information can be calculated as  
\begin{equation}
    K = I_\mathrm{s}(\mathrm{A}:\mathrm{B}) - \chi_\mathrm{E},
\end{equation}
where $I_\mathrm{s}(\mathrm{A}:\mathrm{B})$ denotes the Shannon mutual information between A and B, and $\chi_\mathrm{E}$ represents the eavesdropper's Holevo quantity \cite{Holevo1973, Laudenbach2018}. Figure\,\ref{Fig:Fig5}(b) shows the theoretically expected $K$ as a function of the resource state squeezing level $S$ and noise photon number $n_q$ in the detected quadrature for the case when B performs a homodyne detection on his part of the entangled state. Note that, in contrast to the full noise $n$, we only consider the noise $n_q$, added to the detected quadrature $q$, since homodyne detection is equivalent to phase-sensitive amplification and measurement of a certain quadrature which inherently deamplifies the other quadrature \cite{Adesso2010}. Consequently, only the noise in the amplified quadrature has an effect on the measurement result. We observe that, similarly to the quantities $\Delta_\mathrm{A}$, $\Delta_\mathrm{B}$, and $\Delta_\mathrm{AB}$, the threshold value for $n_q$, when we obtain a positive secret key, converges towards $n_q \simeq 0.26$ for $S \to \infty$. We numerically find that this asymptotic value approximately coincides with $n_*$ implying that, in the high squeezing limit, we can only obtain a positive secret key if the noise, added to the detected quadrature, is lower than the threshold $n_*$, as required for a positive LII flow to the eavesdropper. As a result, the difference between QD and EoF can act as an indicator whether it is possible to obtain a positive secret key or not. More details about the calculation of $K$ are provided in Appendix \ref{Sec:QKD}.

In summary, we have investigated the influence of local noise injection in propagating TMS microwave states on quantum discord correlations and quantum entanglement quantified via the entanglement of formation measure. We have experimentally verified the sudden death of entanglement around theoretically predicted values of approximately one injected noise photon, independent of the squeezing level. Furthermore, we have experimentally demonstrated that in strong contrast to entanglement, QD is asymptotically robust against noise. In addition, we have measured the theoretically predicted crossover between EoF and QD for small noise photon numbers, which is a result of the tripartite nature of mixed TMS states. Since the difference between QD and EoF can be related to the net flow of LII, it may be used to assess the security of certain QKD protocols based on squeezed states. We have demonstrated that the locality of noise injection implies a fundamental difference between the LII flows $\mathrm{A}{\to}\mathrm{B}{\to}\mathrm{E}$ and $\mathrm{B}{\to}\mathrm{A}{\to}\mathrm{E}$. Finally, the demonstrated results on the robustness of QD against noise are relevant for the DQC1 quantum computation approach and quantum illumination protocols. These applications can be viewed as a motivation to intensify the search for quantum information processing, communication, and sensing protocols exploiting QD as a quantum resource. Such protocols would be inherently resistant to noise in contrast to entanglement-based approaches which suffer from the sudden death of entanglement.

We acknowledge support by the German Research Foundation via Germany`s Excellence Strategy (EXC-2111-390814868), the Elite Network of Bavaria through the program ExQM, the EU Flagship project QMiCS (Grant No.~820505), and the German Federal Ministry of Education and Research via the project QUARATE (Grant No.~13N15380), the project QuaMToMe (Grant No.~16KISQ036), JSPS KAKENHI (Grant No.~22H04937), and JST ERATO (Grant No.~JPMJER1601). This research is part of the Munich Quantum Valley, which is supported by the Bavarian state government with funds from the Hightech Agenda Bayern Plus.

\section*{Appendix}
\subsection{Experimental setup}\label{sec:setup}
The JPAs are thermally stabilized at $\SI{50}{\milli \kelvin}$ to guarantee steady squeezing and noise properties. We pump the JPAs using Rohde\&Schwarz SGS100A microwave sources and pulse the pump signal using a data timing generator (DTG) \cite{Zhong_2013, Menzel2012, Pogorzalek2017, Pogorzalek2019}. Here, the TMS state is generated by superimposing two orthogonally squeezed states with equal squeezing levels using a cryogenic hybrid ring, acting as a 50:50 beam splitter. The resulting TMS state at the hybrid ring output then locally looks like thermal noise but shows strong nonlocal correlations in the covariances. The photon number in the TMS state is calibrated using the Planck spectroscopy \cite{Mariantoni2010}, which is realized by sweeping the temperature of a heatable $\SI{30}{\decibel}$ attenuator in the temperature range of $\SI{40}{\milli \kelvin} - \SI{600}{\milli \kelvin}$. By using this heatable attenuator as a self-calibrated cryogenic photon source, we can directly map the detected voltage in the output signal to the photon number in the cryogenic quantum signal \cite{Mariantoni2010}. One part of the TMS state is transmitted to a directional coupler (CPL-4000-8000-20-C, Miteq/Sirius) with coupling $\beta = \SI{-20}{\decibel}$. The coupled port of the directional coupler is used to inject white broadband noise into the system, which is generated using a Keysight 81160A arbitrary waveform generator (AFG). The generated noise has a bandwidth of $\SI{160}{\mega \hertz}$ and is upconverted to the signal reconstruction frequency of $\omega_0/2\pi = \SI{5.323}{\giga \hertz}$ using a local oscillator. Due to the low signal level, the output signal needs to go through multiple amplification stages consisting of a cryogenic high-electron-mobility transistor (HEMT) amplifier and additional room temperature amplifiers, which are stabilized in temperature by a Peltier cooler. The overall noise of the detection chain is determined by the HEMT, which results from its high gain ($\sim \SI{40}{\decibel}$). Frequency-resolved measurements are performed using a vector network analyzer (VNA) and the reconstruction of quantum microwave states as well as quantum correlation measurements are performed using a heterodyne receiver setup. This heterodyne detection setup is similar to the setups described in Refs.\,\citenum{Pogorzalek2017, Pogorzalek2019, Fedorov2018, Fedorov2016}. The signal is down-converted to $\SI{11}{\mega \hertz}$ and digitized by an Acqiris card with a sampling frequency of $\SI{400}{\mega \hertz}$. The digitized data is transmitted to a computer and down converted to a dc-signal. The resulting signal is filtered using a digital finite-impulse-response (FIR) filter with a full bandwidth of $\SI{400}{\kilo \hertz}$. Subsequently, the quadrature moments $\langle I_1^n Q_1^m I_2^k Q_2^l \rangle$, $n, m, k, l \in \mathbb{N}_0$ are determined and averaged. In each measurement, the data is averaged over $210$ cycles, where each cycle corresponds to $5.76 \times 10^6$ averages. The squeezing angles are stabilized using a phase-locked loop, where in each measurement cycle, the squeezing angle $\gamma_i^{\mathrm{exp}}$ is extracted from the quadrature moments corresponding to the $i^{\mathrm{th}}$ JPA, where $i \in \lbrace{ 1, 2 \rbrace}$. Following this approach, the difference $\delta \gamma_i = \gamma_i^{\mathrm{exp}} - \gamma_i^{\mathrm{target}}$ from the desired target angle $\gamma_i^{\mathrm{target}}$ is calculated and the respective phase of the JPA pump source is corrected by $2\delta \gamma_i$. The JPA pump sources are daisy-chained with a reference frequency of $\SI{1}{\giga \hertz}$ to the local oscillator (LO) source. The LO, the DTG, the VNA, the AFG, as well as the Acqiris card are synchronized with a $\SI{10}{\mega \hertz}$ rubidium frequency clock (Stanford Research Systems, FS725). In our experiment, we assume that all reconstructed states are Gaussian and can be described by signal moments up to the second order. This assumption of Gaussianity is verified by calculating the cumulants $\kappa_{mn}$ \cite{Xiang_2018}. Since it is theoretically expected that $\kappa_{mn} = 0$ for $m + n > 2$ for Gaussian states, we conclude that the Gaussian approximation of our quantum states is well justified if the experimentally reconstructed cumulants of third and fourth order are much smaller than the first and second order cumulants. 

\subsection{Quantum key distribution with Gaussian states}\label{Sec:QKD}
In order to investigate the relation between the LII flow and QKD, we consider that Alice and Bob share an ideal TMS state with the squeezing factor $r$, described by the covariance matrix
\begin{equation}
    V_{\mathrm{AB}} = \frac{1}{4}\begin{pmatrix} \cosh \! 2r \, \mathbb{1}_2 & \sinh \! 2r \, \hat{\sigma}_z \\ \sinh \! 2r \, \hat{\sigma}_z & \cosh \! 2r \, \mathbb{1}_2
    \end{pmatrix},
\end{equation}
where $\mathbb{1}_2$ denotes the $2 \times 2$ identity matrix and $\hat{\sigma}_z$ is the Pauli $z$-matrix. For the entangling cloner attack, the eavesdropper prepares a second TMS state \cite{Laudenbach2018}
\begin{equation}
    V_{\mathrm{E}_1\mathrm{E}_2} = \frac{1}{4}\begin{pmatrix} W \mathbb{1}_2 & \sqrt{W^2 - 1} \, \hat{\sigma}_z \\ \sqrt{W^2 - 1} \, \hat{\sigma}_z & W \mathbb{1}_2
    \end{pmatrix}.
\end{equation}
As a result, the full covariance matrix
\begin{equation}
    V_{\mathrm{ABE}_1\mathrm{E}_2} = V_{\mathrm{AB}} \oplus V_{\mathrm{E}_1\mathrm{E}_2}
\end{equation}
describes a pure state. The eavesdropper then couples the mode $E_1$ to B by an asymmetric beam splitter operation
\begin{equation}
    C = \begin{pmatrix}
    \sqrt{1-\beta} \, \mathbb{1}_2 & \sqrt{\beta} \, \mathbb{1}_2 \\
    -\sqrt{\beta} \, \mathbb{1}_2  & \sqrt{1-\beta} \, \mathbb{1}_2
    \end{pmatrix}.
\end{equation}
The resulting covariance matrix is given by
\begin{equation}
    V^\prime_{\mathrm{ABE}_1\mathrm{E}_2} = (\mathbb{1}_2 \oplus C \oplus \mathbb{1}_2) V_{\mathrm{ABE}_1\mathrm{E}_2} (\mathbb{1}_2 \oplus C \oplus \mathbb{1}_2)^\dagger, 
\end{equation}
which can be analytically expressed as
\begin{equation}
    V^\prime_{\mathrm{ABE}_1\mathrm{E}_2} = \frac{1}{4}
    \begin{pmatrix} V^\prime_{11} & V^\prime_{12} & V^\prime_{13} & \mathbb{0}_2 \\ {V^\prime_{12}}^\mathrm{T} & V^\prime_{22} & V^\prime_{23} & V^\prime_{24} \\
    {V^\prime_{13}}^\mathrm{T} & {V^\prime_{23}}^\mathrm{T} & V^\prime_{33} & V^\prime_{34} \\
    \mathbb{0}_2 & {V^\prime_{24}}^\mathrm{T} & {V^\prime_{34}}^\mathrm{T} & V^\prime_{44}
    \end{pmatrix},
\end{equation}
where
\begin{align}\notag
    V^\prime_{11} &= \cosh \! 2r \mathbb{1}_2 \\ \notag
    V^\prime_{12} &= \sqrt{1-\beta} \, \sinh \! 2r \, \hat{\sigma}_z \\ \notag
    V^\prime_{13} &= -\sqrt{\beta} \, \sinh \! 2r \, \hat{\sigma}_z \\ \notag
    V^\prime_{22} &= (1-\beta)\cosh \! 2r \mathbb{1}_2 + \beta W \mathbb{1}_2 \\ \notag
    V^\prime_{23} &= \sqrt{\beta(1-\beta)} \, (W - \cosh \! 2r) \mathbb{1}_2 \\ \notag
    V^\prime_{24} &= \sqrt{\beta} \, \sqrt{W^2 - 1} \, \hat{\sigma}_z \\ \notag
    V^\prime_{33} &= \beta \cosh \! 2r \mathbb{1}_2 + (1-\beta)W \mathbb{1}_2 \\ \notag
    V^\prime_{34} &= \sqrt{1-\beta} \, \sqrt{W^2 - 1} \, \hat{\sigma}_z \\ 
    V^\prime_{44} &= W \mathbb{1}_2.
\end{align}
Since the matrix $V_{22}^\prime/4$ corresponds to the local noisy TMS state in system B, we demand
\begin{equation}
    \frac{1}{4}(\cosh \! 2r + 2n)\mathbb{1}_2  = \frac{1}{4}[(1-\beta)\cosh \! 2r \, \mathbb{1}_2 + \beta W \mathbb{1}_2].
\end{equation}
In the limit $\beta \ll 1$, we find the relation $\beta W = 2n$. Since we perform a homodyne detection on B in the next step, it is practical to define the number of  noise photons, added to the measured quadrature, as $n_q = n/2$. For the final state of the eavesdropper, we have
\begin{equation}
    V_{\mathrm{E}}^\prime = \frac{1}{4}\begin{pmatrix}
    V^\prime_{33}   & V^\prime_{34}  \\
     {V^\prime_{34}}^\mathrm{T}  & V^\prime_{44}
    \end{pmatrix}.
\end{equation}
We consider reverse reconciliation and hence perform a measurement of B \cite{Grosshans2003}. After a homodyne measurement of the $q$ quadrature, the conditioned covariance matrix for E reads
\begin{equation}
V^\prime_{\mathrm{E}|\mathrm{B}} =  V_{\mathrm{E}}^\prime - \frac{1}{4 \sqrt{\det V_{22}^\prime}} V_{\mathrm{C}}^\prime \Pi_q {V_{\mathrm{C}}^\prime}^\mathrm{T}, 
\end{equation}
where $V_{\mathrm{C}} = (V_{23}^\prime, V_{24}^\prime)^\mathrm{T}$ and $\Pi_q$ denotes the phase-space projector on the $q$-quadrature. The corresponding Holevo quantity $\chi_E$ is then obtained as
\begin{equation}
    \chi_\mathrm{E} = S_\mathrm{E} - S_{\mathrm{E}|\mathrm{B}},
\end{equation}
where $S_\mathrm{E}$ ($S_{\mathrm{E}|\mathrm{B}}$) denotes the von Neumann entropy, corresponding to $V_{\mathrm{E}}^\prime$ ($V^\prime_{\mathrm{E}|\mathrm{B}}$) \cite{Holevo1973, Laudenbach2018}. For a two-mode Gaussian state, described by a covariance matrix $V$, the von Neumann entropy is given by
\begin{equation}
    S = f(\nu_+) + f(\nu_-),
\end{equation}
where $\nu_+$ and $\nu_-$ are the symplectic eigenvalues of $V$ \cite{Serafini2003}. For a codebook of input states with variance $\sigma^2$, the Shannon mutual information is obtained to be \cite{Laudenbach2018}
\begin{align}\notag
    I_\mathrm{s}(\mathrm{A}:\mathrm{B}) &= \frac{1}{2}\log_2\,(1 + \mathrm{SNR}) \\ \notag &= \frac{1}{2}\log_2 \! \left(1 + \frac{4(1-\beta)\sigma^2}{(1-\beta)e^{-2r} + 4n_q}\right) \\  &\simeq \frac{1}{2}\log_2 \! \left(1 + \frac{\sigma^2}{n_q}\right),
\end{align}
where the last expression is valid in the limit $r \gg 1$ and $\beta \ll 1$. For the calculation of the signal-to-noise ratio (SNR) in the detected quadrature, we have considered the protocol described in Ref.\,\citenum{Cerf2001}, implying a noise of $e^{-2r}/4 + n_q$ per quadrature and $\sigma^2 = \sinh(2r)/2$. To obtain Fig.\,\ref{Fig:Fig3}, we have fixed $\beta = 10^{-4}$.

\subsection{Fitting the experimental data}\label{Sec:fit}
In this section, we provide details about the model used to fit the experimental data. By taking finite coupling and amplifier noise into account, the final covariance matrix can be expressed as
\begin{equation}\label{Eq: Realistic covariance matrix}
     V_{\mathrm{AB}} = \frac{1 + 2n_\mathrm{j}(G)}{4}\begin{pmatrix}
    \boldsymbol{\alpha} &  \boldsymbol{\gamma} \\
    \boldsymbol{\gamma}^\mathrm{T} &  \boldsymbol{\beta}
    \end{pmatrix},
\end{equation}
where 
\begin{align}
    & \boldsymbol{\alpha} = \cosh \! 2r \, \mathbb{1}_2, \\
    & \boldsymbol{\beta} = \left[(1 - \beta) \cosh \! 2r + \beta (1 + 2 \bar{n}) \right]\mathbb{1}_2, \\
    & \boldsymbol{\gamma} = \sqrt{1 - \beta} \, \sinh \! 2r \, \hat{\sigma}_z.
\end{align}
Furthermore, $n_\mathrm{j}(G)$ represents the noise added by the JPAs, which depends on the degenerate gain $G$. In addition, $\bar{n}$ is the number of noise photons at the input of the coupled port of the directional coupler. The environmental noise $\bar{n}$ is related to $n$ via $n = \beta \bar{n}$ under the assumption that $\bar{n} \gg 1$. The realistic squeezing factor $r$ can be extracted from the reconstructed squeezed (antisqueezed) variance $v_\mathrm{s}$ ($v_\mathrm{a}$) via $e^{4r} = v_\mathrm{a}/v_{\mathrm{s}}$. The degenerate gain can then be expressed as $G = e^{2r}$. Furthermore, we model the gain-dependent JPA noise by a power law dependence $n_\mathrm{j}(G) = \chi_1 (G - 1)^{\chi_2}$, where we treat $\chi_1$ and $\chi_2$ as the only fit parameters. For the fit, we use $\boldsymbol{\chi} \equiv (\chi_1, \chi_2)^\mathrm{T}$ and define a corresponding weighted least-square cost function
\begin{align}
    T(\boldsymbol{\chi}) & = \sum_{S, n} \left(  w_1 \bigg\vert D_\mathrm{A}(S, n, \boldsymbol{\chi}) - \tilde{D}_A(S, n)  \bigg\vert^2 \right. \\ & \left. +   w_2 \bigg\vert D_\mathrm{B}(S, n, \boldsymbol{\chi}) - \tilde{D}_B(S, n)  \bigg\vert^2 \right. \\ & \left. +  w_3 \bigg\vert E_\mathrm{F}(S, n, \boldsymbol{\chi}) - \tilde{E}_\mathrm{F}(S, n)  \bigg\vert^2 \right),
\end{align}
where the sum is evaluated over all experimentally chosen squeezing levels $S$ and noise photon numbers $n$. The quantities $w_i$ are the weights, accounting for the respective contribution. The quantities $\tilde{D}_A(S, n)$, $\tilde{D}_B(S, n)$, and $\tilde{E}_\mathrm{F}(S, n)$ are the experimentally determined data points for A-discord, B-discord, and EoF, respectively, corresponding to $S$ and $n$. The functions $D_\mathrm{A}(S, n, \boldsymbol{\chi})$, $D_\mathrm{B}(S, n, \boldsymbol{\chi})$ and $E_\mathrm{F}(S, n, \boldsymbol{\chi})$ are obtained by inserting Eq.\,\ref{Eq: Realistic covariance matrix} into the theoretical expressions for QD and EoF. The fit parameters $\chi_1 = 0.05$ and $\chi_2 = 0.56$ are then given by
\begin{equation}
    \begin{pmatrix}
         \chi_1 \\
         \chi_2
    \end{pmatrix} = \argmin_{\boldsymbol{\chi}} T(\boldsymbol{\chi}),
\end{equation}
where we start with the initial conditions $\boldsymbol{\chi} = (0, 1)^\mathrm{T}$. To balance the contributions of QD and EoF in the cost function, we choose the weights $w_1 = w_2 = 1/2$ and $w_3 = 1$ for the fit. To extract the noise photon numbers corresponding to the sudden death of entanglement as well as the experimental crossover points between QD and EoF, we do not make use of the fit curves. Instead, we determine these values directly from the experimental data using cubic Hermite spline interpolation as we expect this method to be more accurate than the fit. We use cubic Hermite spline interpolation instead of conventional cubic splines to increase the precision by avoiding overshoots.

\bibliography{Discord}

\end{document}